\documentclass[runningheads]{llncs}

\usepackage{graphicx}
\usepackage{lipsum}
\usepackage{amsmath}
\usepackage{amsfonts}
\usepackage[symbol]{footmisc}
\usepackage{acro}
\usepackage{subcaption}
\usepackage{caption}
\usepackage[table]{xcolor}

\newcommand\blfootnote[1]{%
  \begingroup
  \renewcommand\thefootnote{}\footnote{#1}%
  \addtocounter{footnote}{-1}%
  \endgroup
}

\DeclareMathOperator*{\argmin}{arg\,min}
\DeclareAcronym{SAME}{
short=SAME,
long=SAM-enhanced registration
}
\DeclareAcronym{SAM}{
short=SAM,
long=self-supervised anatomical embedding
}

\DeclareAcronym{SAM-affine}{
short=SAM-affine,
long=SAM-based affine transformation
}

\DeclareAcronym{SAM-coarse}{
short=SAM-coarse,
long=SAM-based coarse deformation
}

\DeclareAcronym{SAM-VM}{
short=SAM-VoxelMorph,
long=SAM-enhanced VoxelMorph
}

\DeclareAcronym{CT}{
short=CT,
long=computed tomography
}
\begin{document}
\title{SAME: Deformable Image Registration based on Self-supervised Anatomical Embeddings}
\titlerunning{SAME: Deformable Image Registration based on Embeddings}
%
\author{Fengze Liu\inst{1}$^\dag$ \and
Ke Yan\inst{2}$^\dag$ \and
Adam Harrison\inst{2} \and
Dazhou Guo\inst{2} \and
Le Lu\inst{2} \and 
Alan Yuille\inst{1} \and
Lingyun	Huang\inst{3} \and 
Guotong	Xie\inst{3} \and 
Jing Xiao\inst{3} \and 
Xianghua Ye\inst{4} \and
Dakai Jin\inst{2}
}
%
\authorrunning{F. Liu et al.}
%
\institute{Johns Hopkins University, Baltimore, MD, USA \and
PAII Inc., Bethesda, MD, USA
\and
Ping An Technology, ShenZhen, China \and 
The First Affiliated Hospital Zhejiang University, Hangzhou, China 
}

\maketitle              
\begin{abstract}

In this work, we introduce a fast and accurate method for unsupervised 3D medical image registration. This work is built on top of a recent algorithm \ac{SAM}, which is capable of computing dense anatomical/semantic correspondences between two images at the pixel level. Our method is named \ac{SAME}, which breaks down image registration into three steps: affine transformation, coarse deformation, and deep deformable registration. Using \ac{SAM} embeddings, we enhance these steps by finding more coherent correspondences, and providing features and a loss function with better semantic guidance. We collect a multi-phase chest computed tomography dataset with 35 annotated organs for each patient and conduct inter-subject registration for quantitative evaluation. Results show that \acs{SAME} outperforms widely-used traditional registration techniques (Elastix FFD, ANTs SyN) and learning based VoxelMorph method by at least $4.7\%$ and $2.7\%$ in Dice scores for two separate tasks of within-contrast-phase and across-contrast-phase registration, respectively. SAME achieves the comparable performance to the best traditional registration method, DEEDS (from our evaluation), while being orders of magnitude faster (from 45 seconds to 1.2 seconds).

\keywords{Deformable Registration \and Affine Registration \and Unsupervised \and Self-supervised Anatomical Embedding \and Deep Learning.}
\end{abstract}

\blfootnote{$^\dag$ equal contribution.}

\acresetall
\section{Introduction}

Deformable image registration is a fundamental task in medical image analysis~\cite{Rueckert2011}. Traditional registration methods solve an optimization problem and iteratively minimize a preset similarity measure to align a pair of images. Recently, learning-based deformable registration, using deep networks, have been  investigated~\cite{Balakrishnan2019VM,Dual-Stream,Recursive,Pyramid,Liu_2020}. Compared with their conventional counterparts, learning-based methods can incorporate more flexible losses, integrate other computing modules and are much faster in inference. VoxelMorph was a representative work~\cite{Balakrishnan2019VM} that learns a parameterized registration function using a convolutional neural network (CNN). Many recent methods focus on designing more sophisticated networks using pyramid~\cite{Pyramid} or cascaded structures~\cite{Dual-Stream,Recursive}, or connecting registration to pipelines that include synthesis and segmentation~\cite{Liu_2020}. Ideally, registration should focus on aligning semantically similar/coherent voxels, e.g., the same anatomical locations. This semantic information can come in the form of extra manual annotations (e.g.~organ masks)~\cite{Balakrishnan2019VM}, but requiring prohibitive labor costs from professionals. Existing unsupervised methods instead optimize similarity measures describing local intensities as a proxy of the semantic information, such as the mean squared error (MSE) or normalized cross correlation (NCC). However, these are less reliable in settings with large deformations, complex anatomical differences, or cross-modality/cross-phase imagery. 

In this paper, we exploit incorporating a novel form of semantic information in registration. \Ac{SAM} is a recent work as a means to produce pixel-wise embeddings in radiological images by encoding anatomical semantic information~\cite{Yan2020SAM}. 
It requires no annotations in training. SAM can match corresponding points between two images, which is exactly the fundamental goal of image registration. 
The most simple and straightforward way to register two images with SAM is to extract SAM embeddings from both fixed and moving images, match each moving pixel to the closest fixed pixel in SAM space, and calculate the corresponding coordinate offsets to generate a deformation field. However, this approach is highly inefficient, as there are millions of pixels in a typical 3D \ac{CT} scan. Besides, SAM would not incorporate spatial smoothness constraints~\cite{Balakrishnan2019VM}, which is useful when the correspondences predicted by SAM contain noises.

We propose \acf{SAME} to address these issues. \ac{SAME} is comprised of three consecutive steps. (1) \textbf{\acs{SAM-affine}}, which uses correspondence points generated from SAM on a sparse grid to compute the affine transformation matrix. Affine registration~\cite{Klein2010Elastix} has been widely used either alone or as an initialization of deformable methods~\cite{Balakrishnan2019VM,Heinrich2012DEEDS}. (2) \textbf{\acs{SAM-coarse}}, which uses a coarse correspondence grid to directly produce a coarse-level deformation field. These first two steps are efficient, require no additional training, and can provide a good initialization for the final step. (3) Lastly, \textbf{\acs{SAM-VM}} enhances the deep learning-based  VoxelMorph registration method~\cite{Balakrishnan2019VM}, using SAM-based correlation features~\cite{Dosovitskiy2015FlowNet} and a newly formulated SAM similarity loss. \acs{SAME} is evaluated on a multi-phase chest CT dataset for inter-subject registration with $35$ thoracic organs annotated. Quantitative experimental results show that \acs{SAM-affine} significantly outperforms traditional optimization-based affine registration in both accuracy and speed. The complete \acs{SAME} consistently outperforms traditional approaches~\cite{Rueckert1999FFD,Avants2008ANTS} and VoxelMorph~\cite{Balakrishnan2019VM} in both within-contrast-phase and across-contrast-phase tasks by average Dice scores of $4.7\%$ and $2.7\%$, respectively. SAME matches DEEDS~\cite{Heinrich2012DEEDS}, as the state-of-the-art in CT registration~\cite{Xu_2016}, while being orders of magnitude faster (1.2 sec vs.~45 sec).

\section{Method}
In this section, we present the details of the proposed \acs{SAME} for deformable registration 
and describe how \acs{SAM} is integrated in each of the three steps.

\subsection{\Acf{SAM}} 
SAM is recently proposed by~\cite{Yan2020SAM}, as a novel pixel-level contrastive learning framework with a coarse-to-fine network and a hard-and-diverse negative sampling strategy. In an unsupervised manner, it predicts a global and a local embedding vector with semantic meanings per pixel in a CT volume---the same anatomical location in different images expressing similar embeddings. \Acs{SAM} is readily used to find correspondences between images, providing a means to solve the registration problem from a new perspective. Let $X_{f},X_{m} \in \mathbb{R}^{D \times H \times W}$ be the fixed and moving images to be registered
. For each image, we extract the global and local SAM embedding volumes and concatenate them in the channel dimension, resulting in $S_f, S_m \in \mathbb{R}^{C \times D \times H \times W}$ ($C$ is the concatenated channel dimension). Given a point $p_f = (x, y, z)$ in $X_f$, we take its embedding vector $S_f(:, z, y, x)$ and convolve it with $S_m$ to get a similarity heatmap volume. The point with the highest similarity score becomes the matched point in the moving image. Results show that matching for a single point only consumes $0.2$ sec on a common chest CT scan~\cite{Yan2020SAM}. 

\begin{figure}[t]
\includegraphics[width=\textwidth]{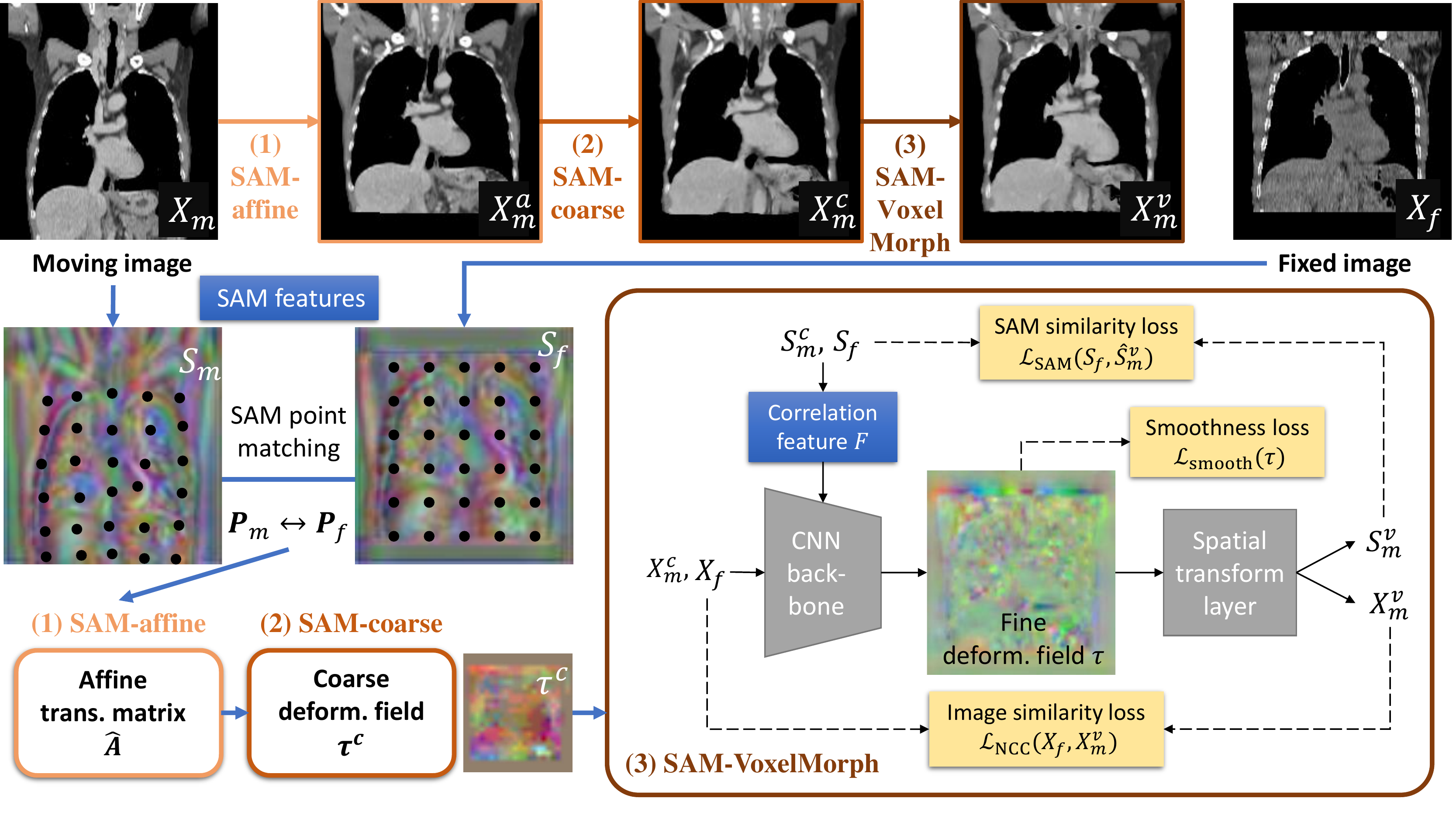}
\caption{\acf{SAME} framework. The moving image is warped by three consecutive steps: \acs{SAM-affine}, \acs{SAM-coarse}, \acs{SAM-VM}, gradually approaching the fixed image. Variables $X$, $S$, and $P$ denote the image, SAM embedding, and point coordinates, respectively. Subscripts $m$, $f$ stand for moving or fixed, respectively. Superscripts $a, c$ and $v$ indicate the variable is generated after each of the three steps (affine, coarse deform, or VoxelMorph).} \label{framework} 
\end{figure}

\subsection{\Ac{SAM-affine} and \Ac{SAM-coarse}} 

Matched SAM correspondences can be directly employed to estimate an affine transformation matrix~\cite{Klein2010Elastix,Heinrich2012DEEDS,Balakrishnan2019VM}. First, we select a set of points on $X_f$ for matching. Intuitively, evenly distributed points on the image may lead to a better estimation. Therefore, we use the points on a regular grid on $X_f$, see Fig.~\ref{framework}. It would be more precise to run point matching on every pixel (instead of a coarse grid) and directly generate a fine deformation field, but that would consume $0.5$h for a CT with 200 slices. To balance accuracy and speed, we use a grid with stride $8$. Since SAM is only designed for points inside the body, we segment the body mask of $X_f$ using intensity thresholding and morphological post processing, and then remove grid points outside the mask. When doing point matching, we downsample $S_m$ with spatial stride of $4$ to reduce computation. After the corresponding points in $X_m$ are located, we need to filter out low-quality matches. We examine their similarity scores and discard those lower than a threshold $\theta$. After that, we can get $k$ matched points in $X_f, X_m$, which can be represented by $3\times k$ matrices: $\mathbf{P}_f$ and $\mathbf{P}_m$, respectively.   We pad them with 1s to create homogeneous versions of the matched points coordinates, $\tilde{\mathbf{P}}_f$, $\tilde{\mathbf{P}}_m \in \mathbb{R}^{4\times k}$, and estimate the affine matrix $\hat{\mathbf{A}} \in \mathbb{R}^{4\times 4}$ by a simple least squares fitting: 
\begin{align} \label{least_squares}
    \hat{\mathbf{A}}=\argmin_{\mathbf{A}}{\|\mathbf{A}\tilde{\mathbf{P}}_m-\tilde{\mathbf{P}}_{f}\|_F^2}.
\end{align}

Next, we transform $X_m$ with $\hat{\mathbf{A}}$ to obtain $X_m^a$ 
and extract new SAM embeddings $S_m^a$ from it. Then, points in $\mathbf{P}_f$ are matched again on $X_m^a$ to get $\mathbf{P}_m^a$. $\mathbf{P}_m^a$ and $\mathbf{P}_f$ actually represent a mapping from $X_m^a$ to $X_f$ on $k$ sparse points. We can compute their difference $\Delta = \mathbf{P}_f - \mathbf{P}_m^a$, and map each point in $\Delta$ back to the original coordinates of the image to get $\tau^c \in \mathbb{R}^{3 \times D \times H \times W}$. Note, there are only $k$ deformation in $\Delta$ that are not necessarily uniformly spaced. Thus values in  $\tau^c$ are filled in using linear interpolation. This gives us the final coarsely estimated deformation map, which is applied to warp $(X_m^a, S_m^a)$ to $(X_m^c, S_m^c)$
.  Although coarsely estimated (on only $k$ points), $\tau_c$ can effectively reduce the difference between the moving and the fixed images. Compared to a global affine alignment, this provides local warps that can serve as a better initialization for a final learning-based deformable registration step. One question is that whether we could omit SAM-affine and compute $\tau^c$ directly. We observed that before affine registration, the two images may have significant offsets, so $\tau^c$ is potentially large in magnitude, which will magnify the noises in the matched points. Thus, we first perform affine registration to reduce the magnitude of deformations.

\subsection{\Ac{SAM-VM}} 

The objective of the final step is to predict a fine deformation map $\tau \in \mathbb{R}^{3 \times D \times H \times W}$, which is a spatial transformation function that can warp the moving image to best match the fixed one. Following the framework of VoxelMorph~\cite{Balakrishnan2019VM}, we learn a function $\Phi:(X_{f},X_{m}^c)\rightarrow \tau$ with a CNN
. The original VoxelMorph uses pure pixel intensity-based features and similarity losses. We improve them by leveraging the semantic information contained in \acs{SAM} embeddings using SAM correlation features and a SAM loss (see Fig.~\ref{framework}).

The loss function in VoxelMorph and follow-up works includes two parts, an image similarity loss and a smoothness loss. We use the local normalized cross-correlation (NCC) loss~\cite{Balakrishnan2019VM} for the former,
while the latter is defined as 
\begin{align}
\mathcal{L}_{smooth}(\tau)=\frac{1}{|\Omega|}\sum_{\textbf{u}\in\Omega}||\nabla\tau_{\textbf{u}}||^2 \mathrm{,}
\end{align}
where $\Omega$ is the set of all pixels within the body mask. However, the NCC loss only compares local image intensities, which may not be robust under CT contrast injection, pathological changes, and large or complex deformations in the two images. On the other hand, the SAM embeddings can uncover semantic similarities between two pixels. 
Thus, we add a proposed SAM loss:
\begin{align}
\mathcal{L}_{SAM}(S_f,S_m^v)=\frac{1}{|\Omega|}\sum_{\textbf{u}\in\Omega}\langle S_f(\textbf{u}),S_m^v(\textbf{u}) \rangle \mathrm{,}
\end{align}
where the superscript $v$ indicates the feature map has been warped by $\tau$ predicted by \acs{SAM-VM}. The final loss is
\begin{equation} \label{final_loss}
\mathcal{L} = \mathcal{L}_{NCC}(X_f,X_m^v) +\lambda \mathcal{L}_{SAM}(S_f,S_m^v)+\gamma \mathcal{L}_{smooth}(\tau).
\end{equation}

While the SAM loss is an effective means to more semantically align images, the \emph{features} extracted in standard VoxelMorph still lack semantic information, which may be needed to better guide predictions. The correlation feature was originally proposed in FlowNet~\cite{Dosovitskiy2015FlowNet} to manage this problem for optical flow. It was also used in \cite{Heinrich2020highly} for registration. Briefly, it computes the similarity of pixel $\textbf{u}$ on $X_f$ and pixel $\textbf{u}+\textbf{d}$ on $X_m$, where $\textbf{d}$ is a small displacement. This similarity is computed for each pixel and for $n$ possible displacement values to generate an $n$-channel feature map, which is then concatenated to the original feature map at some point in the network. 
When using SAM, the semantic similarity of two pixels can be simply computed as the inner product of two SAM vectors, $F(\textbf{u})=\langle S_{f}(\textbf{u}),S_{m}^c(\textbf{u}+\textbf{d}) \rangle$. We empirically find that using $27$ displacement values $\textbf{d} \in \{-2, 0, 2\}^3$ yields good results. Injecting the SAM correlation features provides improved cues to the network when predicting deformations, thus brings further boosts in accuracy.




\begin{table}[tbp]

    \centering
 \caption{Comparison of different registration methods. We show the average Dice score (\%) of two tasks: CE-to-CE and CE-to-NC registration. VM: VoxelMorph. Best and second best performance is shown in bold and gray box, respectively.}
    \label{tab:registration}
    \setlength{\tabcolsep}{6pt}
    \begin{tabular}{c c c c c}
\hline

\multicolumn{1}{l}{Methods}  & \multicolumn{1}{|c}{CE-to-CE} &
\multicolumn{1}{|c}{CE-to-NC} &  \multicolumn{1}{|p{1.5cm}}{Inference time (s)} & 
\multicolumn{1}{|c}{std of $|J_{\phi}|$}\\
\hline
\multicolumn{1}{l}{Elastix-affine~\cite{Klein2010Elastix}}  & \multicolumn{1}{|c}{28.44} &
\multicolumn{1}{|c}{27.96} &  \multicolumn{1}{|c}{3.38} &  \multicolumn{1}{|c}{-}\\
\multicolumn{1}{l}{MIND-affine~\cite{Heinrich2012MIND}}  & \multicolumn{1}{|c}{28.24} &
\multicolumn{1}{|c}{27.91} &  \multicolumn{1}{|c}{7.86} &  \multicolumn{1}{|c}{-}\\ 
\multicolumn{1}{l}{SAM-affine (SA)}  & \multicolumn{1}{|c}{33.80} &
\multicolumn{1}{|c}{33.77} &  \multicolumn{1}{|c}{0.48}&  \multicolumn{1}{|c}{-} \\
\multicolumn{1}{l}{SAM-coarse (SC)}  & \multicolumn{1}{|c}{44.67} &
\multicolumn{1}{|c}{43.68} &  \multicolumn{1}{|c}{0.78}&  \multicolumn{1}{|c}{-} \\
\multicolumn{1}{l}{SA + SC}  & \multicolumn{1}{|c}{46.76} &
\multicolumn{1}{|c}{45.67} &  \multicolumn{1}{|c}{1.05}&  \multicolumn{1}{|c}{0.40} \\
\hline
\multicolumn{1}{l}{SA +  VM~\cite{Balakrishnan2019VM}}  & \multicolumn{1}{|c}{48.79} &
\multicolumn{1}{|c}{47.35} &  \multicolumn{1}{|c}{0.78}&  \multicolumn{1}{|c}{0.38} \\
\multicolumn{1}{l}{SA + SAM-VM}  & \multicolumn{1}{|c}{51.99} &
\multicolumn{1}{|c}{49.90} &  \multicolumn{1}{|c}{0.84}&  \multicolumn{1}{|c}{0.36} \\
\multicolumn{1}{l}{SA + SC +  VM}  & \multicolumn{1}{|c}{\cellcolor[HTML]{EFEFEF}54.12} &
\multicolumn{1}{|c}{50.64} &  \multicolumn{1}{|c}{1.13} &  \multicolumn{1}{|c}{0.68}\\
\multicolumn{1}{l}{SA + SC + SAM-VM (ours)}  & \multicolumn{1}{|c}{\textbf{54.42}} &
\multicolumn{1}{|c}{\cellcolor[HTML]{EFEFEF}50.96} &  \multicolumn{1}{|c}{1.16} &  \multicolumn{1}{|c}{0.66}\\

\hline
\multicolumn{1}{l}{SyN~\cite{Avants2008ANTS}}  & \multicolumn{1}{|c}{49.75} &
\multicolumn{1}{|c}{47.95} &  \multicolumn{1}{|c}{74.34} &  \multicolumn{1}{|c}{-}\\
\multicolumn{1}{l}{FFD~\cite{Rueckert1999FFD}}  & \multicolumn{1}{|c}{49.36} &
\multicolumn{1}{|c}{48.22} &  \multicolumn{1}{|c}{93.51}&  \multicolumn{1}{|c}{0.51} \\
\multicolumn{1}{l}{DEEDS~\cite{Heinrich2012DEEDS}}  & \multicolumn{1}{|c}{52.72} &
\multicolumn{1}{|c}{\textbf{51.15}} &  \multicolumn{1}{|c}{45.35} &  \multicolumn{1}{|c}{0.40}\\

\hline

\end{tabular}
{\scriptsize *Paired t-tests show SAME significantly outperforms all other methods ($p<10^{-4}$), except for DEEDS in the CE-to-NC setting. SAM-VM significantly outperforms VM ($p<10^{-7}$).}

{\scriptsize **The average surface distance (ASD) in CE-to-CE: FFD 4.6mm, SA+VM 4.1mm, DEEDS 4.0mm, SA+SAM-VM 3.9mm, SA + SC + SAM-VM 3.8mm.}
\label{registration}
\end{table}
\section{Experiments}  

\subsubsection{Dataset and task.} To evaluate \ac{SAME}, we collected a chest CT dataset containing $94$ subjects, each with a contrast-enhanced (CE) and a non-contrast (NC) scan. We randomly split the patients to $74$, $10$, and $10$ for training,  validation, and testing. Each image has manually labeled masks of $35$ organs (including lung, heart, airway, esophagus, aorta, bones, muscles, arteries and veins)~\cite{Guo2021LNS}. For the validation and test sets, we construct $90$ image pairs for inter-subject registration and calculate an atlas-based segmentation accuracy on the $35$ organs. Performances of two tasks are evaluated: intra-phase registration (CE-to-CE) and cross-phase registration (CE-to-NC). Every image is resampled to an isotropic resolution of $2$mm and cropped to $208 \times 144 \times 192$ by clipping black borders. The image intensity is normalized to $(-1,1)$ using a window of $(-800, 400)$ HU.

\subsubsection{Implementation details.} Our method was developed using PyTorch 1.5. It was run on a Ubuntu server with 12 CPU cores of 3.60GHz. It requires one NVIDIA Quadro RTX 6000 GPU to train and test. We trained a SAM model using the training set of the chest CT dataset. Its structure is identical with the one in~\cite{Yan2020SAM}, which outputs a 128D global embedding and a 128D local one for each pixel. This model is fixed and applied in all three steps of SAME. In \acs{SAM-affine} and \acs{SAM-coarse}, the similarity threshold $\theta$ is set to 0.7 to select high-confidence matches. In \acs{SAM-VM}, we use a 3D progressive holistically-nested network (P-HNN)~\cite{Harrison2017PHNN} as the backbone and concatenate the correlation feature before the third convolutional block. We also tried 3D U-Net~\cite{Cicek2016Unet} but observed no significant accuracy gains. The loss weights in Eq.~\ref{final_loss} are empirically set to $\lambda=1, \gamma=0.5$. We train \acs{SAM-VM} using the Adam optimizer with a learning rate of 0.001 for 10 epochs. Each training batch contains 2 image pairs with random contrast phases (CE or NC). We evaluate the registration results using average Dice score over $35$ organ masks. The organ masks are not used during training.

\subsubsection{Quantitative results.}
From Table~\ref{registration} we can see that \textbf{SAM-affine} outperforms the traditional affine registration method in Elastix~\cite{Klein2010Elastix} by 5-6\%, meanwhile being 6 times faster. It is also better than affine registration with the MIND~\cite{Heinrich2012MIND} robust descriptor. This is because SAM can match corresponding anatomical locations between two images accurately and efficiently. Compared with other methods that iteratively optimizes the affine parameters, SAM-affine directly calculates affine matrix by least squared fitting. 
\textbf{SAM-coarse} surpasses SAM-affine by $10\%$ since it allows for locally deformable warping with  more degrees of freedom. Cascading these two steps further boosts the accuracy. VoxelMorph pre-aligned by SAM-affine outperforms SAM-affine + SAM-coarse moderately since the latter can only perform a coarse deformable transformation. However, note that the former is a learning-based dense registration method, while the latter does not require any extra training. It only utilizes the matching result of a pretrained SAM model on grid points. The $2\%$ small gap demonstrates the capability of our proposed SAM-coarse.

\begin{table}[tbp]

    \centering
 \caption{Ablation study for different settings on incorporating SAM to VoxelMorph (VM). The average Dice score (\%) is reported. All methods are initialized by SAM-affine without SAM-coarse.}
    \label{tab:ablation}
    \setlength{\tabcolsep}{6pt}
    \begin{tabular}{c c c c c }
\hline
\multicolumn{1}{c}{Methods}  & \multicolumn{1}{|c}{SAM loss} &
\multicolumn{1}{|c}{SAM correlation feature} & \multicolumn{1}{|c}{CE-to-CE} & \multicolumn{1}{|c}{CE-to-NC} \\

\hline
\multicolumn{1}{c}{VM~\cite{Balakrishnan2019VM}}  & \multicolumn{1}{|c}{$\times$} &
\multicolumn{1}{|c}{$\times$} & \multicolumn{1}{|c}{48.79} & \multicolumn{1}{|c}{47.35} \\
\hline
\multicolumn{1}{c}{}  & \multicolumn{1}{|c}{\checkmark} &
\multicolumn{1}{|c}{$\times$} & \multicolumn{1}{|c}{50.43} & \multicolumn{1}{|c}{48.24} \\
\multicolumn{1}{c}{SAM-VM}  & \multicolumn{1}{|c}{$\times$} &
\multicolumn{1}{|c}{\checkmark} & \multicolumn{1}{|c}{51.37} & \multicolumn{1}{|c}{48.99} \\
\multicolumn{1}{c}{}  & \multicolumn{1}{|c}{\checkmark} &
\multicolumn{1}{|c}{\checkmark} & \multicolumn{1}{|c}{\textbf{51.99}} & \multicolumn{1}{|c}{\textbf{49.90}} \\
\hline

\hline
\end{tabular}
\label{ablation} 
\end{table}

\begin{figure}[t]
\includegraphics[width=.96\textwidth]{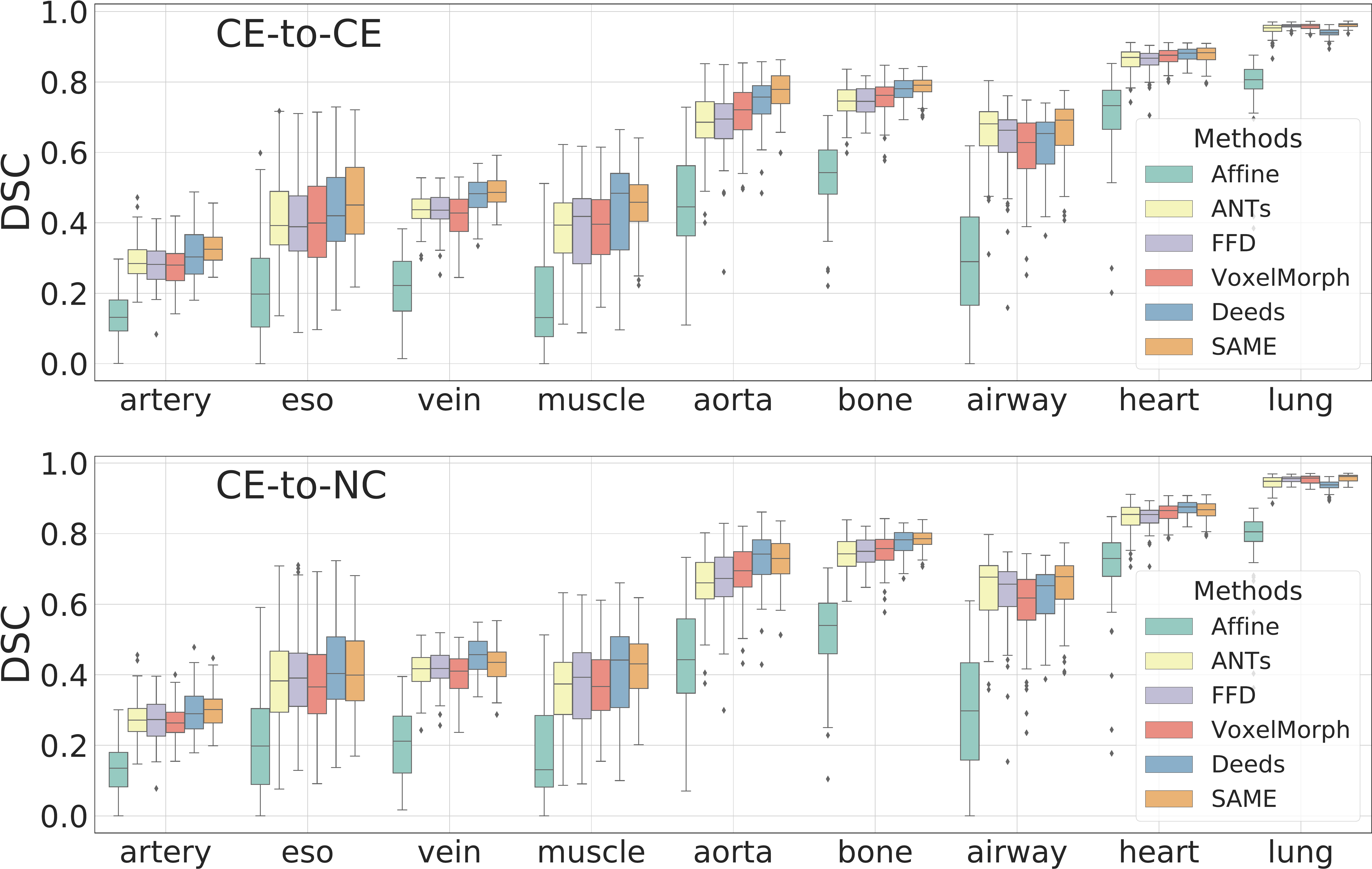}
\caption{Comparison of registration methods on all organ groups. Eso: esophagus.}
\label{fig:compare}
\end{figure}
\begin{figure}[!t]
\includegraphics[width=1\textwidth]{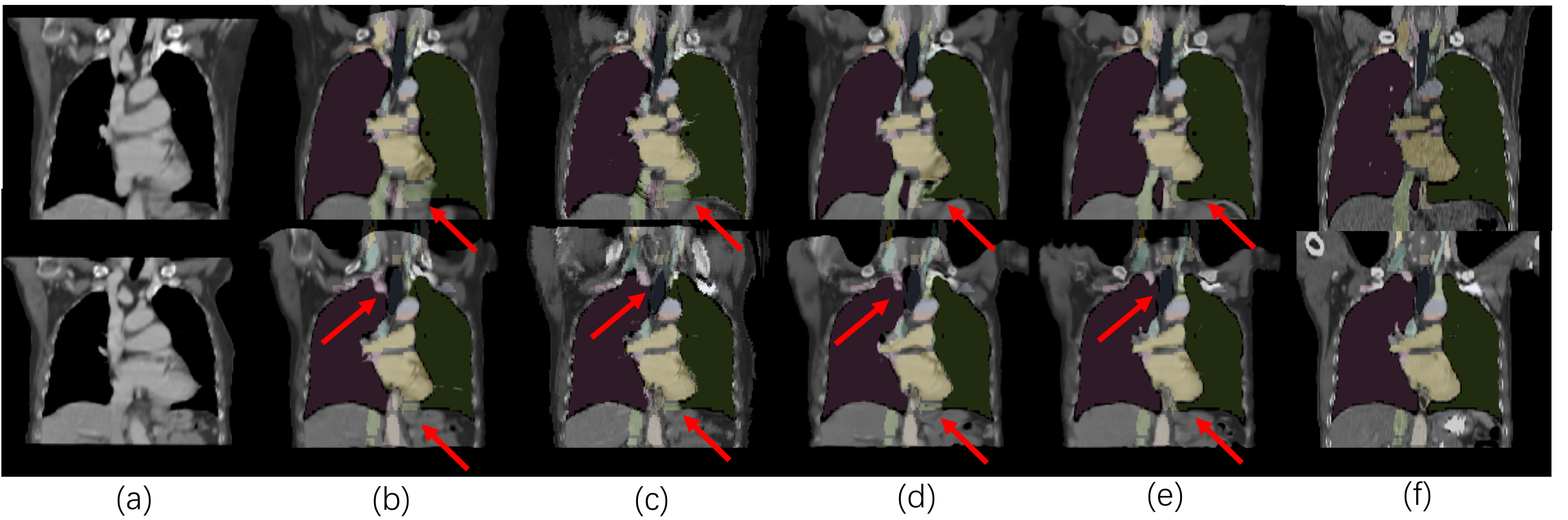}
\caption{Visualization of registration results from different methods. From left to right is (a) the moving image, (b) warped moving image of ANTs, (c) DEEDS, (d) SAM-affine + VoxelMorph, (e) SAME, and (f) the fixed image.}
\label{fig:example} 
\end{figure}

SAM-affine + SAM-coarse can provide a good initialization to the learning-based VM in the third step, allowing it to better perform. From the 4 rows in the middle block of Table~\ref{registration}, we also observe consistent improvement by replacing the original VoxelMorph~\cite{Balakrishnan2019VM} with \textbf{SAM-VM}. The SAM embeddings contain more semantic information than the raw pixel intensities, which is incorporated to SAM-VM by the SAM-based correlation feature and SAM loss. 
An ablation study of SAM-VM is shown in Table~\ref{ablation}, where the best result is achieved when both the correlation feature and SAM loss are used. 
On one hand, explicitly inputting the correlation feature calculated by SAM provides extra guidance for determining the deformation fields
. On the other hand, 
the SAM loss provides a more semantically informed supervisory signal. 

In the bottom block of Table~\ref{tab:registration}, we evaluate several widely-used non-rigid registration methods including FFD~\cite{Rueckert1999FFD}, SyN~\cite{Avants2008ANTS}, and DEEDS~\cite{Heinrich2012DEEDS}. FFD was implemented using Elastix [10], where parameters matched the best performing FFD method in EMPIRE10 Challenge~\cite{murphy2011evaluation}. The only modification was an extra bending energy term with weight 0.01 to regularize the smoothness. For SyN (implemented in ANTS) and DEEDS (implemented by the original author), parameters were set according to those used in~\cite{Xu_2016}. For affine transform, the default implementation in each package was used. The proposed SAME (combination of three steps) achieves markedly better results than SyN and FFD. Compared with the best traditional method (DEEDS), it performs better in the within-phase setting and comparably in the cross-phase setting, meanwhile is 38 times faster.  Cross-phase registration is more difficult because the brightness and appearance of contrast-enhanced and non-contrast CTs can be very different (see $X_m$ and $X_f$ in Fig.~\ref{framework}), and DEEDS has explicitly designed the modality independent features in its registration. \ac{SAME} takes a different approach that uses the modality invariant SAM embeddings to align images. 

We have computed the standard deviation of Jacobian determinants to measure the smoothness of the deformation field. In Table~\ref{tab:registration}, it is observed that SAME achieves the best Dice with a certain degree of sacrifice in smoothness. This is mainly because SAME cascades two deformable methods, SAM-coarse (SC) and SAM-VM. The smoothness of SAM-VM alone is slightly better than the original VM (0.36 vs. 0.38), but SC itself brings more non-smoothness (0.40). SC generates a deformation field by directly differentiating two sets of coordinates without any constraint. This approach gives SC more flexibility to model large deformation but may also produce less smoothed results. We will study on adding constraints to improve the smoothness of SC in the future. On the other hand, if SC is not used, SA + SAM-VM can also achieve competing accuracy (52.0\% Dice score) with good smoothness (0.36), where the overall performance is still comparable to DEEDS (52.7\%, 0.40) while significantly better than FFD (49.4\%, 0.51), and SA+VM (50.8\%, 0.38).

Organ-specific results are shown in Fig.~\ref{fig:compare}. For the sake of conciseness, we divide the $35$ organs in our dataset into $9$ groups and calculate the median and inter-quartile range of Dice score within each group. The affine in Fig.~\ref{fig:compare} is from Elastix~\cite{Klein2010Elastix}, whereas the VoxelMorph refers to SAM-affine + VM~\cite{Balakrishnan2019VM} in Table~\ref{registration}. The results of SAME surpass DEEDS on 8 out of 9 groups except heart in the within-phase condition. In the cross-phase setting, SAME outperforms DEEDS on the artery, bone, airway and lung organs. In other organs, like esophagus and muscle, SAME shows results with smaller variance and comparable median performance with DEEDS. Organ groups such as artery, esophagus, vein, and muscle display lower Dice scores for all methods because they are typically small and can be confused with surrounding tissues. Qualitative examples are illustrated in Fig.~\ref{fig:example}. Manual organ masks of the fixed images are overlaid to show whether the warped moving images align well with the fixed image. Arrows pointed to regions where SAME works better than other methods.

\section{Conclusion}

In this paper, we propose SAME, a fast and accurate framework for unsupervised medical image registration. We expect SAM-affine and SAM-coarse to be promising alternatives of traditional optimization-based methods for registration initialization. The SAM correlation feature and SAM loss may also be combined with other learning-based algorithms~\cite{Liu_2020,Recursive} for further accuracy improvement.
%
%
%
\bibliographystyle{splncs04}
\bibliography{main}

\begin{thebibliography}{10}
\providecommand{\url}[1]{\texttt{#1}}
\providecommand{\urlprefix}{URL }
\providecommand{\doi}[1]{https://doi.org/#1}

\bibitem{Avants2008ANTS}
Avants, B.B., Epstein, C.L., Grossman, M., Gee, J.C.: {Symmetric diffeomorphic
  image registration with cross-correlation: Evaluating automated labeling of
  elderly and neurodegenerative brain}. Med. Image Anal.  \textbf{12}(1),
  26--41 (2008). \doi{10.1016/j.media.2007.06.004}, \url{www.itk.org}

\bibitem{Balakrishnan2019VM}
Balakrishnan, G., Zhao, A., Sabuncu, M.R., Guttag, J., Dalca, A.V.:
  {VoxelMorph: A Learning Framework for Deformable Medical Image Registration}.
  IEEE Transactions on Medical Imaging  \textbf{38}(8),  1788--1800 (2019).
  \doi{10.1109/TMI.2019.2897538}, \url{http://voxelmorph.csail.mit.edu.}

\bibitem{Cicek2016Unet}
{\c{C}}i{\c{c}}ek, {\"{O}}., Abdulkadir, A., Lienkamp, S.S., Brox, T.,
  Ronneberger, O.: {3D U-net: Learning dense volumetric segmentation from
  sparse annotation}. In: MICCAI. vol. 9901 LNCS, pp. 424--432 (2016)

\bibitem{Dosovitskiy2015FlowNet}
Dosovitskiy, A., Fischery, P., Ilg, E., Hausser, P., Hazirbas, C., Golkov, V.,
  Smagt, P.V.D., Cremers, D., Brox, T.: {FlowNet: Learning optical flow with
  convolutional networks}. In: ICCV. vol. 2015 Inter, pp. 2758--2766 (2015).
  \doi{10.1109/ICCV.2015.316}

\bibitem{Guo2021LNS}
Guo, D., Ye, X., Ge, J., Di, X., Lu, L., Huang, L., Xie, G., Xiao, J., Lu, Z.,
  Peng, L., Yan, S., Jin, D.: {DeepStationing: Thoracic Lymph Node Station
  Parsing in CT Scans using Anatomical Context Encoding and Key Organ
  Auto-Search }. In: MICCAI. vol.~LNCS (2021)

\bibitem{Harrison2017PHNN}
Harrison, A.P., Xu, Z., George, K., Lu, L., Summers, R.M., Mollura, D.J.:
  {Progressive and multi-path holistically nested neural networks for
  pathological lung segmentation from CT images}. In: MICCAI. vol. 10435 LNCS
  (2017), \url{https://adampharrison.gitlab.io/p-hnn/}

\bibitem{Heinrich2020highly}
Heinrich, M.P., Hansen, L.: {Highly accurate and memory efficient unsupervised
  learning-based discrete CT registration using 2.5 D displacement search}. In:
  MICCAI (2020)

\bibitem{Heinrich2012MIND}
Heinrich, M.P., Jenkinson, M., Bhushan, M., Matin, T., Gleeson, F.V., Brady,
  S.M., Schnabel, J.A.: {MIND: Modality independent neighbourhood descriptor
  for multi-modal deformable registration}. Medical Image Analysis
  \textbf{16}(7),  1423--1435 (2012). \doi{10.1016/j.media.2012.05.008},
  \url{http://users.ox.ac.uk/{~}shil3388/}

\bibitem{Heinrich2012DEEDS}
Heinrich, M.P., Jenkinson, M., Brady, S.M., Schnabel, J.A.: {Globally optimal
  deformable registration on a minimum spanning tree using dense displacement
  sampling}. In: MICCAI. vol. 7512 LNCS, pp. 115--122 (2012)

\bibitem{Dual-Stream}
Hu, X., Kang, M., Huang, W., Scott, M.R., Wiest, R., Reyes, M.: Dual-stream
  pyramid registration network. In: Shen, D., Liu, T., Peters, T.M., Staib,
  L.H., Essert, C., Zhou, S., Yap, P.T., Khan, A. (eds.) Medical Image
  Computing and Computer Assisted Intervention -- MICCAI 2019. Springer
  International Publishing (2019)

\bibitem{Klein2010Elastix}
Klein, S., Staring, M., Murphy, K., Viergever, M.A., Pluim, J.P.: {Elastix: A
  toolbox for intensity-based medical image registration}. IEEE Transactions on
  Medical Imaging  \textbf{29}(1),  196--205 (2010).
  \doi{10.1109/TMI.2009.2035616}, \url{http://elastix.isi.uu.nl/wiki.php}

\bibitem{Liu_2020}
Liu, F., Cai, J., Huo, Y., Cheng, C.T., Raju, A., Jin, D., Xiao, J., Yuille,
  A., Lu, L., Liao, C., Harrison, A.P.: Jssr: A joint synthesis, segmentation,
  and registration system for 3d multi-modal image alignment of large-scale
  pathological ct scans. In: Vedaldi, A., Bischof, H., Brox, T., Frahm, J.M.
  (eds.) Computer Vision -- ECCV 2020. pp. 257--274. Springer International
  Publishing, Cham (2020)

\bibitem{Pyramid}
Mok, T.C.W., Chung, A.C.S.: Large deformation image registration with
  anatomy-aware laplacian pyramid networks. Segmentation, Classification, and
  Registration of Multi-modality Medical Imaging Data: MICCAI 2020 Challenges,
  ABCs 2020, L2R 2020, TN-SCUI 2020, Held in Conjunction with MICCAI 2020,
  Lima, Peru, October 4, 2020, Proceedings  \textbf{12587},  61--67 (Feb 2021)

\bibitem{murphy2011evaluation}
Murphy, K., Van~Ginneken, B., Reinhardt, J.M., Kabus, S., Ding, K., Deng, X.,
  Cao, K., Du, K., Christensen, G.E., Garcia, V., et~al.: Evaluation of
  registration methods on thoracic ct: the empire10 challenge. IEEE
  transactions on medical imaging  \textbf{30}(11),  1901--1920 (2011)

\bibitem{Rueckert1999FFD}
Rueckert, D., Sonoda, L.I., Hayes, C., Hill, D.L.G., Leach, M.O., Hawkes, D.J.:
  {Nonrigid Registration Using Free-Form Deformations: Application to Breast MR
  Images}. IEEE Trans. Med. Imaging  \textbf{18}(8) (1999)

\bibitem{Rueckert2011}
Rueckert, D., Schnabel, J.A.: Medical Image Registration, pp. 131--154.
  Springer Berlin Heidelberg, Berlin, Heidelberg (2011)

\bibitem{Xu_2016}
{Xu}, Z., {Lee}, C.P., {Heinrich}, M.P., {Modat}, M., {Rueckert}, D.,
  {Ourselin}, S., {Abramson}, R.G., {Landman}, B.A.: Evaluation of six
  registration methods for the human abdomen on clinically acquired ct. IEEE
  Transactions on Biomedical Engineering  \textbf{63}(8),  1563--1572 (2016)

\bibitem{Yan2020SAM}
Yan, K., Cai, J., Jin, D., Miao, S., Harrison, A.P., Guo, D., Tang, Y., Xiao,
  J., Lu, J., Lu, L.: {Self-supervised learning of pixel-wise anatomical
  embeddings in radiological images} (2020),
  \url{https://arxiv.org/abs/2012.02383}

\bibitem{Recursive}
{Zhao}, S., {Dong}, Y., {Chang}, E., {Xu}, Y.: Recursive cascaded networks for
  unsupervised medical image registration. In: 2019 IEEE/CVF International
  Conference on Computer Vision (ICCV). pp. 10599--10609 (2019).
  \doi{10.1109/ICCV.2019.01070}

\end{thebibliography}
\end{document}